\documentclass{article}
\usepackage[utf8]{inputenc}     
\usepackage[english]{babel}     
\usepackage[a4paper, left=1.5in, right=1.5in, top=1.5in, bottom=1.5in]{geometry}                              
\usepackage[x11names,dvipsnames,svgnames]{xcolor}
\usepackage{graphicx}           
\usepackage{amsmath,amssymb}    
\usepackage{amsthm}             
\usepackage{mathtools}   
\usepackage{lastpage}
\usepackage{enumitem}  
\usepackage{lipsum}
\usepackage{setspace}
\usepackage{fancyhdr}
\usepackage{setspace}
\usepackage{listings}   
\usepackage{titlesec}
\usepackage{titletoc}
\usepackage{caption}
\usepackage{zref-totpages}
\usepackage[colorinlistoftodos]{todonotes}
\usepackage{subcaption}
\usepackage{epigraph}
\usepackage{orcidlink}
\usepackage{hyperref} 
\hypersetup{
    colorlinks,
    citecolor=blue,
    filecolor=blue,
    linkcolor=blue,
    urlcolor=blue
}
\usepackage{listings}
\usepackage[T1]{fontenc}
\usepackage{titlesec, color}
\definecolor{gray75}{gray}{0.75}

\titleformat{\section}{\Large\bfseries}{\thesection.\ }{0pt}{\Large\bfseries}
\titleformat{\subsection}{\large\bfseries}{\thesubsection.\ }{0pt}{\large\bfseries}

\usepackage{natbib}
\setcitestyle{aysep={}}

\fancypagestyle{plain}{
  \fancyhf{}
  \fancyhead[C]{\small Imagination and Understanding through Astrophysical Imagery}
  \fancyfoot[R]{\small page \thepage\ of \ztotpages\hspace{4mm}}

}

\fancypagestyle{mystyle}{
  \fancyhf{}
  \fancyhead[C]{\small Imagination and Understanding through Astrophysical Imagery}
  \fancyfoot[R]{\small page \thepage\ of \ztotpages\hspace{4mm}}

}

\fancypagestyle{front}{
  \fancyhf{}
  \fancyfoot[R]{\small page \thepage\ of \ztotpages\hspace{4mm}}

}

\begin{document}

\thispagestyle{front}
\setstretch{1.2}
\begin{center}
\huge\textbf{Imagination and Understanding through Astrophysical Imagery}\\
\vspace{7mm}
\Large Paul\ H.W.\ Disberg \orcidlink{0000-0002-0492-4089}\\
\end{center}
\vspace{2mm}
\footnotesize \textbf{--} Center for the History of Philosophy and Science, Radboud University, Nijmegen, The Netherlands\\
\textbf{--} Department of Astrophysics / IMAPP, Radboud University, Nijmegen, The Netherlands\\
\textbf{--} School of Physics and Astronomy, Monash University, Clayton, Australia\\
\textbf{--} The ARC Center of Excellence for Gravitational Wave Discovery---OzGrav, Australia\vspace{-2mm}\\
\\
\footnotesize Contact:\ \href{mailto:paul.disberg@monash.edu}{paul.disberg@monash.edu}\vspace{2mm}\\
\footnotesize Final version accepted for publication in \href{https://direct.mit.edu/posc}{Perspectives on Science} (MIT Press)\vspace{2mm}\\
\footnotesize Submitted 14 August 2024 / Accepted 22 March 2025
\vfill
\section*{Abstract}
\normalsize Scientific articles, for instance in the field of astrophysics, are often filled with a variety of images. In philosophical studies, these images are usually analyzed in terms of their function within the scientific argument presented in the article. However, not all images that can be found in astrophysical articles are relevant to the scientific argument, which prompts the question of why they are included in the first place. Using the example of the so-called \textquotedblleft Stellar Graveyard\textquotedblright\ plot, I argue that the work of Letitia Meynell provides a valuable description of this kind of imagery. That is, there are images used in astrophysical literature that may not be necessary for the scientific argument, but function as an aide for the visual imagination of the reader. These kinds of aides can help with mentally visualizing certain spatial configurations and the causal relationships within them, ultimately furthering understanding of the discussed astrophysical concepts or models.
\phantomsection
\addcontentsline{toc}{section}{Acknowledgments}
\section*{Acknowledgments}
I am deeply thankful to Christoph Lüthy and Henk de Regt for supervising the Master's thesis that formed the basis for this article and for providing extremely valuable comments and feedback, based on which the article was substantially improved. I am also grateful to the anonymous referees for elaborate comments which helped increase the quality of the argumentation and structure of present article. I thank Frederik Bakker and Nicola Gaspari for useful discussions, and Gijs Nelemans, Andrew Levan, and Ilya Mandel for their guidance in the world of astrophysics. Moreover, I am thankful to the authors and institutions that---where relevant---granted permission to use their figures. Lastly, I acknowledge support from the Australian Research Council Centre of Excellence for Gravitational Wave Discovery (OzGrav) through project number CE230100016.
\newpage
\pagestyle{mystyle}

\section{Introduction}
\label{sec1}
\epigraph{\textquotedblleft Imagination is more important than knowledge.\\ Knowledge is limited.\\ Imagination encircles the world.\textquotedblright}{Albert Einstein \citep[in][p.\ 117]{Viereck_1929}}
\noindent Images play an important role in all scientific fields: scientists use images to communicate their findings in papers, present them at conferences, or use them in lectures. When walking through a department of science at a university, one finds the walls often filled with posters of research projects, all containing images representing the results. Moreover, when reading an article, scientists scan for images in order to assess its content and determine its relevance, making the images an important tool of communication. In fact, the process of writing a scientific paper often starts with representing the research in figures, after which the text is added to bind the figures together into a coherent argument. For this reason, I am specifically interested in the images labeled as \textquotedblleft figures\textquotedblright\ in contemporary scientific papers. These include graphs, which show a numerical relation between quantities, and diagrams, which do not.\footnote{See \citet{Luthy_2009} for a more elaborate discussion of these concepts.\vspace{-5mm}}
\par
\begin{figure}
    \centering 
    \includegraphics[width=\linewidth]{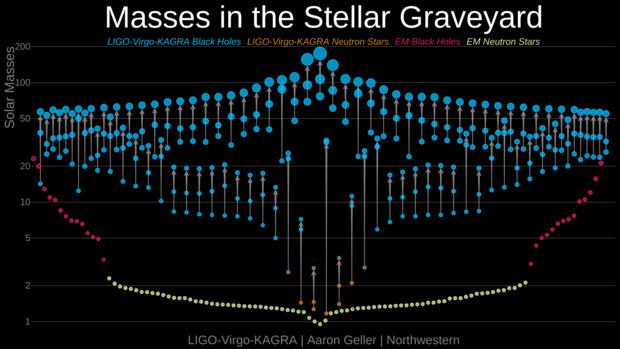}
    \caption{The Stellar Graveyard plot, displaying the observed neutron stars (yellow and orange) and black holes (red and blue). The horizontal axis is an arbitrary ordering, while the vertical axis shows the masses of the objects. The blue dots show the black holes found in binaries, where two smaller black holes merge into a bigger one. The name \textquotedblleft graveyard\textquotedblright\ refers to the fact that neutron stars and black holes are stellar remnants, and could therefore be called \textquotedblleft dead stars.\textquotedblright\ Image credit: \href{https://media.ligo.northwestern.edu/gallery/mass-plot}{LIGO-Virgo-KAGRA / Aaron Geller / Northwestern University}.}
    \label{fig1}
\end{figure}

 This article is motivated by one scientific figure in particular: the so-called \textquotedblleft Stellar Graveyard\textquotedblright\ plot, shown in Figure \ref{fig1}. This plot displays the known neutron stars and black holes, which are either observed through light (EM) or through gravitational waves (by LIGO-Virgo-KAGRA). A gravitational wave is a distortion in space that travels at the speed of light and is created when two extremely dense objects, such as black holes, orbit each other, spiral inwards (due to energy loss because of the emission of gravitational waves), and eventually merge into one, more massive, black hole. The shape of the gravitational wave, then, depends on the masses of the two merging black holes, enabling astrophysicists to estimate the masses of the two black holes that orbit each other (and thus are in a \textquotedblleft binary\textquotedblright).
\par The Stellar Graveyard plot (Figure \ref{fig1}) displays the masses of these black holes, as blue (solid) circles, where the arrows show how the two binary black holes merge into a more massive black hole (i.e., the two smaller blue circles are connected to a bigger blue circle). This figure is used often when discussing the population of black hole binaries: in department talks, conferences, lectures and in articles \citep[e.g.,][]{Miller_2019,Bodensteiner_2022,Broadhurst_2022,Mandel_2022,Waldrop_2022}. One is almost guaranteed to encounter this image when doing research into the population of binary black holes, which means that there must be something which makes the Stellar Graveyard plot an attractive image.
\par
\begin{figure}
    \centering 
    \includegraphics[width=\linewidth]{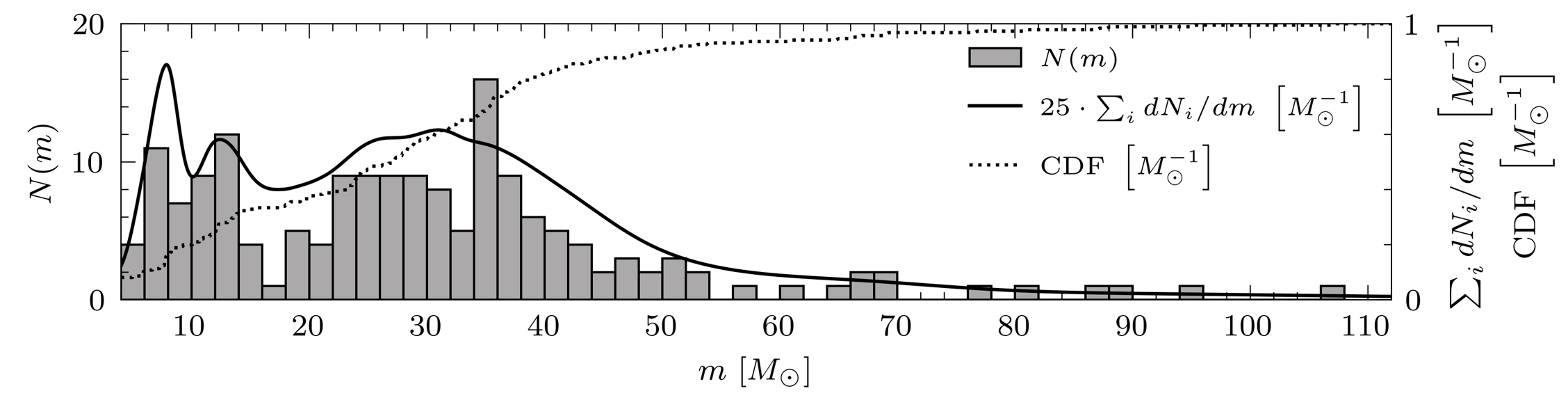}
    \caption{Binary black hole mass distribution as observed through gravitational waves \citep[amended version of Fig.\ 1 of][reproduced following the CC BY 4.0 license]{Disberg_2023}. The black holes which are present as blue circles in Figure \ref{fig1} are here represented through several statistical tools: a histogram (gray squares), a sum of Gaussian distributions (solid line), and a cumulative distribution function (dotted line).}
    \label{fig2}
\end{figure}

 However, this image is a considerably sub-optimal way of presenting the masses of these black holes. The overall distribution of masses is difficult to infer from this image, because the black holes in Figure \ref{fig1} are shown as individual circles, and the x-axis has no physical meaning at all \citep[\textquotedblleft Placement on the horizontal axis is arbitrary,\textquotedblright][p.\ 3]{Mandel_2022}. For scientific purposes, a representation such as the one shown in Figure \ref{fig2} is far better at showing the mass distribution. In this figure, a histogram shows the number of times a certain mass occurs in the distribution, instead of representing the masses individually. Aspects relevant for astrophysical research become clear, such as the apparent gap at $m\approx17M_{\odot}$ \citep[see also, e.g.,][]{Schneider_2023,Adamcewicz_2024}, which is not detectable in Figure \ref{fig1}. This is why technical papers on the binary black hole mass distribution often contain a representation similar to Figure \ref{fig2}.
\par When the Stellar Graveyard figure is used in articles, the authors might use it as a way to show the typical masses of binary black holes. For example, they might state that the black holes observed through gravitational waves are more massive than the ones seen through electromagnetic waves (i.e., light): \textquotedblleft Thus, the LIGO detections immediately doubled the mass range of known stellar-mass black holes, and then the merger doubled the range again (see [their version of the Stellar Graveyard plot])!\textquotedblright\ \citep[][p.\ 471]{Miller_2019}, and \textquotedblleft [T]he black holes found by the Laser Interferometer Gravitational-Wave Observatory and its European counterpart (LIGO/Virgo) span a much larger mass range (3–100 $M_{\odot}$; see [Figure \ref{fig1}])\textquotedblright\ \citep[][p.\ 4]{Bodensteiner_2022}. However, since a figure similar to Figure \ref{fig2} is better at showing the typical masses of the black holes and a better tool for a comparison between the populations, why is the Stellar Graveyard plot used here? Moreover, sometimes the Stellar Graveyard is present in a paper without supporting any specific statement the authors make in the text. For example, \citet{Mandel_2022} simply refer to the Stellar Graveyard figure in their article as: \textquotedblleft The masses of the black-hole binaries observed to date are shown in [Figure \ref{fig1}], along with masses inferred from X-ray binary observations.\textquotedblright\ \citep[][p.\ 3]{Mandel_2022}. That is, they merely use the Stellar Graveyard to display the masses of the black holes, even though a figure such as Figure \ref{fig2} harbors more information regarding the masses of the black holes.
\par This prompts the question: if the representation of the black hole masses as shown in the Stellar Graveyard plot is sub-optimal for scientific purposes, then why is this plot so popular? In order to answer this question, I will mainly turn to the work of Letitia \citet{Meynell_2017,Meynell_2018,Meynell_2020} to argue that (1) the Stellar Graveyard plot is attractive, not because of its representation of the scientific data, but because it appeals to the imagination, (2) this imaginative function is strongly connected with scientific understanding, and (3) imagery that exhibits this function is prevalent in astrophysical articles. That is, I will investigate in what way imagery, imagination, and understanding relate to each other in the context of astrophysics / science, and whether astrophysical literature shows more examples of images exhibiting this relation.
\par In Section \ref{sec2}, I will discuss three philosophical accounts of the functions of scientific imagery that can shed a light on several aspects of the Stellar Graveyard image. That is, I will compare accounts that describe images as propositional claims (by Laura Perini), as models or descriptions of models (by Stephen Downes), or as aides for the imagination (by Letitia Meynell). Based on these accounts, I will argue that the Stellar Graveyard is a good example of an image meant to aid the visual imagination, as described by Meynell. Subsequently, in Section \ref{sec3}, I will discuss the role that the scientific imagination can play in astrophysical research, for example with the purpose of furthering understanding, and give more example of astrophysical imagery that in my estimation also exhibit the function described by Meynell. Finally, I will summarize my conclusions in Section \ref{sec4}.
\newpage
\section{Functions of Scientific Imagery}
\label{sec2}
It goes beyond the scope of this paper to give a comprehensive overview of the philosophical literature on scientific imagery, but in order to examine the role scientific images such as the Stellar Graveyard can play in scientific articles, I will consider two philosophical accounts that, although they might describe elements of the figure, appear to be insufficient to describe the function of the image. In particular, I will discuss the work of Laura \citet{Perini_2005,Perini_2005b,Perini_2012} who describes images as truth-bearers (Section \ref{sec2.1}), and the work of Stephen \citet{Downes_2009,Downes_2011,Downes_2012} who describes scientific images as representative models (Section \ref{sec2.2}). Then, I will discuss the philosophical account of Letitia \citet{Meynell_2017,Meynell_2018,Meynell_2020} who establishes the relationship between imagery and the imagination (Section \ref{sec2.3}). I will argue that, in contrast to the first two accounts, Meynell's account seems to provide a satisfying description of the Stellar Graveyard plot (Section \ref{sec2.4}).
\subsection{Propositional Images}
\label{sec2.1}
One way a scientific image can function is to \textquotedblleft bear truth,\textquotedblright according to \citet{Perini_2005,Perini_2005b,Perini_2012}. In other words, this role of scientific imagery entails representing the world to support claims that can either be true or false. The argument Perini makes in favor of this position builds on a logical distinction between linguistic and pictorial symbol systems, using the concepts introduced by \citet{Goodman_1976}. One of the important differences mentioned by Perini is that in linguistic systems the individual symbols (e.g., letters) are clearly distinguishable, while this is not the case in pictorial systems. For example, if the $y$-values of a graph are shifted by a small value $\delta$, the difference would be clearly identifiable in an equation describing the graph. However, if $\delta$ is smaller than the size of a pixel, the (pictorial) graphs would be indistinguishable.
\par \citet{Perini_2005b} discusses the capacity of pictures to have a truth-value. She argues that pictures are a representation of a certain \textquotedblleft state of affairs,\textquotedblright\ which means these pictures can provide support for linguistic statements describing this state. Using the example of a picture made through electron microscopy (i.e., a \textquotedblleft micrograph\textquotedblright), she states:
\begin{quote}
    \textquotedblleft The micrograph is a representation of a structural configuration of stained material. A pictorial representation of that state of affairs provides support for statements that are linguistic claims about the fact represented by the figure. If the state of affairs represented by the micrograph obtains, then a linguistic description of some aspect of that state of affairs is true. (...) So the first step in reasoning from the micrograph is an inference that draws a linguistically expressed conclusion about the structure of the sample from the micrograph — a pictorial representation of the structure of the sample.\textquotedblright\ \citep[][p.\ 920]{Perini_2005b}.
\end{quote}
For Perini, it is important that the linguistic statement (e.g., \textquotedblleft the particles are uniformly distributed over the sample\textquotedblright) is being inferred from the picture, because this implies that the picture is the fundamental truth-value laden representation. Also, her usage of the term \textquotedblleft state of affairs\textquotedblright\ implies she uses a correspondence theory of truth. This means that a statement, or a picture in this case, is true if it corresponds to the actual world. Similarly, the Stellar Graveyard plot can be said to be true or false, depending on whether it accurately shows the masses of the black holes as observed through gravitational waves. That is, it is possible to infer linguistic claims (which have a truth-value) from the \textquotedblleft mathematical structure\textquotedblright\ \citep[][p.\ 279]{Perini_2005} of pictorial systems such as diagrams, graphs, and pictures, for instance through conventions about their meaning, such as a linear axis representing time.\footnote{See \citet{Luthy_2024} for a historical analysis.\vspace{-5mm}} Perini concludes that there is no reason to reserve the role of truth-bearing for linguistic systems. In other words, she states that pictorial systems such as the ones described above can in fact \textquotedblleft bear truth.\textquotedblright\ After all, pictures can most certainly correspond to the actual world, in which case they are \textquotedblleft true\textquotedblright\ according to the correspondence theory of truth. 
\par The motivation for Perini's analysis is the assumption that (1) figures play a genuine role in scientific arguments and (2) the only way this can happen is if figures have a truth-value. Especially the second assumption has been criticized \citep[e.g.,][]{Goodwin_2009}. However, \citet{Perini_2012} reacts to this by turning to scientific practice and stressing the role images play in supporting the conclusion of a paper:
\begin{quote}
    \textquotedblleft  Authors worry about whether they are correct or not, whether or not their conclusion is conveyed through an equation, a statement, or a diagram. Referees evaluate whether such conclusions are adequately supported by the evidence. When evidence is presented in diagrammatic format, referees evaluate whether it adequately supports the conclusion of the paper.\textquotedblright\ \citep[][p.\ 146]{Perini_2012}.
\end{quote}
This means that, for Perini, the main function of the claims that follow from this kind of scientific imagery---which are either true or false---is to support the conclusion of the scientific argument. Therefore, scientists must aim to design these images to support their claims optimally, in order to provide the maximum support for the conclusion. The strict evaluations by editors and referees help to ensure this goal is achieved. Perini views scientific articles as making an argument in favor of this conclusion:
\begin{quote}
    \textquotedblleft When scientists introduce and defend hypotheses, they do what philosophers do: give talks and publish articles. These presentations amount to arguments: they are supposed to provide justification for belief in new, even controversial, ideas.\textquotedblright\ \citep[][p.\ 262]{Perini_2005}.
\end{quote}
The scientific arguments she describes can be viewed as being logical: there are premises which lead to a conclusion. To summarize the relevant properties of images that fulfill this function: propositional images (1) support a claim which can be either true or false, (2) play a relevant role in the scientific argument, and (3) are intended to provide optimal support for their claims.
\par Figure \ref{fig2} is a good example of a propositional image: it supports the claim that the masses of the black holes observed through gravitational waves are as depicted, and would be \textquotedblleft false\textquotedblright\ if it did not correspond to the actual observations. Moreover, it plays a significant role in the scientific argument of \citet{Disberg_2023}, who aimed to explain the gap at $m\approx17M_{\odot}$, and it is therefore designed to optimally display the mass distribution and its apparent gap. Examples of propositional images in astrophysics can include graphs inferred from simulations \citep[e.g., Figure 4 of][]{Disberg_2025} or observations \citep[e.g., Figure 1 of][]{Arcavi_2017}, or sometimes even pictures of the observations themselves \citep[e.g., Figure 2 of][]{Levan_2024}.
\subsection{Images as Models}
\label{sec2.2}
An alternative function of scientific imagery is described by \citet{Downes_2009,Downes_2011,Downes_2012}, who first of all argues that scientific images play diverse roles in science. Because of this, he argues, philosophers will be unlikely to formulate a theory of scientific imagery that attributes one specific functionality to all scientific images. Downes acknowledges that some images might function as truth-bearers (as described by Perini), but also points at images for which this may not be an accurate description. As an example of images that have no propositional function, he makes a comparison between images and models:
\begin{quote}
    \textquotedblleft What some scientific images have in common with most scientific models is that they are taken by scientists to stand for some physical system or pattern in a data set.\textquotedblright\ \citep[][p.\ 420]{Downes_2009}.
\end{quote}
With \textquotedblleft to stand for\textquotedblright\ Downes refers to a similarity relation between model and object. He builds on the philosophy of \citet{Giere_1988}, who views models and images as being similar in \textquotedblleft specific respects and degrees\textquotedblright\ to reality \citep[][p.\ 421]{Downes_2009}. This can entail resembling the spatial structure of an (astrophysical) object, but Downes also mentions Bohr's atom model as an example of a model that resembles the real object in a more abstract-theoretical way, while still being useful for scientific analysis.
\par \citet{Downes_2011} refers to this notion (i.e., the function of models as resembling some real object) and notes that the resemblance between the model and the world can have a certain degree of accuracy:
\begin{quote}
    \textquotedblleft On a view like this, models stand in some representational relation to a world system or observable aspects of that system. This view comes along with a system of epistemic appraisal centered around the fit of the model with the world or the observed world. Rather than being true or false, models are claimed to be isomorphic with the system they represent, or similar to that system in various respects and degrees.\textquotedblright\ \citep[][p.\ 759]{Downes_2011}.
\end{quote}
This means that a model cannot be said to be true or false\footnote{As statistician George \citet[][]{Box_1979} said: \textquotedblleft All models are wrong but some are useful.\textquotedblright\ \citep[][p.\ 2]{Box_1979}.\vspace{-5mm}} and a diagrammatic image of a model can therefore not be called a propositional image. That is, models should not be viewed as either true or false, but as being accurate to a certain degree, and therefore useful within a given context. A significant degree of similarity between model and reality, then, makes the model more accurate and can provide justification for a conclusion based on that model, resulting in \textquotedblleft epistemic appraisal\textquotedblright\ (as described by Downes). How well the model fits reality (or rather \textquotedblleft observation,\textquotedblright\ as scientists of course do not have direct access to reality) provides a measure of success for the scientific analysis:
\begin{quote}
    \textquotedblleft On this account science is successful to the extent that scientists produce models of the world that are similar to their objects, or observable aspects of those objects, in relevant respects and degrees.\textquotedblright\ \citep[][p.\ 760]{Downes_2011}.
\end{quote}
An image that functions as a model, then, is useful to scientists to the extent that it is an accurate representation of the object of interest.
\par \citet{Downes_2012} does not only argue that scientific images can be \textquotedblleft descriptions\textquotedblright\ of models, but also that they can be models themselves. That is, not only can images display \textquotedblleft idealizations and assumptions\textquotedblright\ \citep[][p.\ 128]{Downes_2012} that make up a certain model, but according to Downes images can also themselves be the model that is used within the scientific analysis:
\begin{quote}
    \textquotedblleft When the relevant image is what scientists use to further inquiry into an area and the image does not stand for or abbreviate a set of equations or clearly describe or specify another model, I take the image to be a model.\textquotedblright\ \citep[][p.\ 128]{Downes_2012}.
\end{quote}
Moreover, Downes stresses that an image functioning as a (description of a) model can represent a \textquotedblleft highly abstract\textquotedblright\ scientific idea, meaning there is no necessary truth-value laden correspondence between the image and the outside world and scientific imagery is therefore not necessarily propositional.
\begin{figure}
    \centering 
    \includegraphics[width=.5\linewidth]{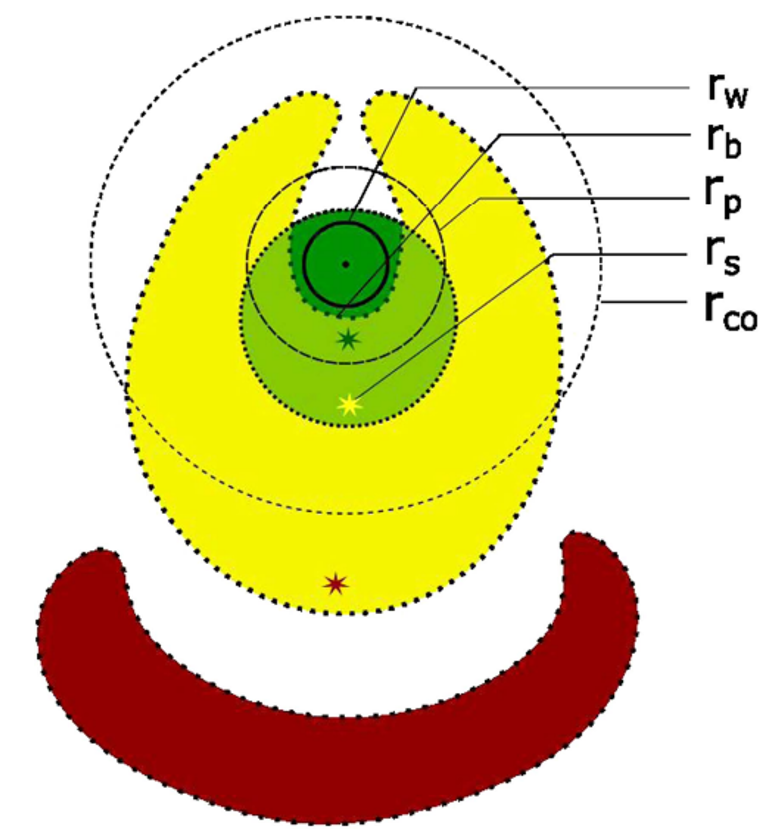}
    \caption{Model of supernova winds \citep[Fig.\ 1 of][\textcopyright\,AAS---reproduced with permission]{Yalinewich_2017}. The image shows the possible structures of matter expelled in a supernova, depending on its position relative to the Galactic center.}
    \label{fig3}
\end{figure}
\par Figure \ref{fig3} shows an image from a scientific article whose function seems to be accurately described by Downes' account. It shows matter expelled from a supernova, depending on how close to the Galactic center the supernova is located. The several distances marked in the diagram, such as the maximum distance for the supernova matter to be able to engulf the Galactic center, are important for the scientific analysis by \citet{Yalinewich_2017}. In their article, the authors refer to the image, and the three scenarios it depicts, as follows:
\begin{quote}
    \textquotedblleft For a typical supernova explosion with $E\approx10^{51}$ erg and mass $M_e\approx5M_{\odot}$, explosions occurring at distances below $2$ pc will penetrate the wind-generating zone. The morphology of such explosions is shown as a green circle in [Figure \ref{fig3}], and its simulation is discussed in Section 3.2. Explosions at distances between $2$ and $50$ pc will engulf the wind-generating zone, but will not penetrate it. This morphology is shown as a yellow horseshoe in [Figure \ref{fig3}], and its simulation is discussed in Section 3.3. At distances larger than $50$ pc the explosion will be overwhelmed. This morphology is shown as a red crescent in [Figure \ref{fig3}], and its simulation is discussed in Section [...].\textquotedblright\ \citep[][pp.\ 7--8]{Yalinewich_2017}.
\end{quote}
This shows that the image indeed displays the different morphologies that the scientists are interested in recreating through simulation, and effectively forms a hypothesis for the eventual results of the simulations. This image is not propositional: it is difficult to consider the displayed morphologies to have a truth-value. Instead, Figure \ref{fig3} represents the general shape of the different morphologies and the abstract relationship between these morphologies and the distance of the supernova to the Galactic center. These distances are defined in the figure and explicitly used in the calculations made by the authors, which can be viewed as describing the assumptions that go into their model. The model (description) is important for the scientific analysis performed by the authors. I argue that the main function of the figure is to visualize these model elements and the image is therefore well-described by the account of Downes.
\subsection{Imaginative Images}
\label{sec2.3}
Letitia \citet{Meynell_2017,Meynell_2018,Meynell_2020}, in contrast to Perini and Downes, argues that there are scientific images whose purpose does not appear to be the accurate representation of things such as real-world objects. Instead, she identifies images that seem to be aimed at appealing to the imagination of their audience. Usually, philosophers distinguish between two types of imagination: propositional and imagistic imagination. \citet{McGinn_2004}, for instance, calls these two types \textquotedblleft imagining-that\textquotedblright\ and \textquotedblleft mental images,\textquotedblright\ respectively. He illustrates this difference by referring to an argument by Descartes, regarding the ability to visualize a polygon with a thousand sides (i.e., a chiliogon):
\begin{quote}
    \textquotedblleft [I]magery is not the same as imagining-that. I can imagine \textit{that} there is a chiliogon in the next room; I just cannot form an image of one (that is distinct from my image of a 1,001-sided figure). I can combine the concepts in question and adopt the attitude of imagination toward the resulting propositional content; I just cannot make my \textit{image} that fine-grained (...).\textquotedblright\ \citep[][pp.\ 128--129, italics in original]{McGinn_2004}.
\end{quote}
Imagining-that is therefore not necessarily the same as forming a mental image, and McGinn calls the latter \textquotedblleft mindsight,\textquotedblright\ contrasting it with visualization through perception. \citet{Rocke_2010} applies the theory of McGinn to scientific practice, and refers to the concept of \textquotedblleft mindsight\textquotedblright\  as \textquotedblleft image-ination\textquotedblright.\footnote{Cf.\ the term \textquotedblleft IMAGination,\textquotedblright\ as used by \citet[][p.\ 312]{Miller_1984}.\vspace{-5mm}} He defines this as: \textquotedblleft the ability to form and manipulate mental images that are not directly connected to visual perception itself\textquotedblright\ \citep[][p.\ 324]{Rocke_2010}, in order to distinguish between, for instance, a mental image of a chair formed by observing a chair (visual perception), and one formed by imagining a chair (image-ination). In her review of literature concerning imagination in science, \citet{Murphy_2022} briefly addresses the role imagery can play in helping the imagination. She states that images can indeed \textquotedblleft aid the imagination\textquotedblright: 
\begin{quote}
    \textquotedblleft Furthermore, scientists often make use of diagrams and pictures when presenting models, thought experiments and other scientific representations (...). These function as aids to our imagination – they help to direct our attention in the right kinds of way, enabling us to grasp the key features of the described setup more readily.\textquotedblright\ \citep[][p.\ 5]{Murphy_2022}.
\end{quote}
That is, authors can use an image to guide the image-ination of their audience. Murphy refers to the work of \citet{Meynell_2017}, who gives a more elaborate discussion of this process within the context of thought experiments. Meynell argues there is \textquotedblleft an important interaction between images and [thought experiments] that rests on their crucially sharing a certain imaginative character\textquotedblright\ \citep[][p.\ 509]{Meynell_2017}. 
\par In order to describe how actual images can evoke mental ones, Meynell builds on the theory of \citet{Walton_1990}, who discusses representation in general and states that a proposition can be fictionally true, or \textquotedblleft true in a game of make-believe\textquotedblright\ \citep[][p.\ 35]{Walton_1990}. This means that there can be a fictional world in which certain statements could be considered to be \textquotedblleft true.\textquotedblright\ An example of such a fictional world might be children pretending that tree branches are swords. According to Walton, certain statements, such as \textquotedblleft The children are holding swords,\textquotedblright\ can be assigned a truth-value within the fictional world of the game. Subsequently, he introduces the concept of \textquotedblleft prop,\textquotedblright\ which is an attribute used in generating the fictional world:
\begin{quote}
    \textquotedblleft Props are generators of fictional truths, things which, by virtue of their nature or existence, make propositions fictional. A snow fort is a prop. It is responsible for the fictionality of the proposition that there is a (real) fort with turrets and a moat. (...) Props are often prompters or objects of imagining also; (...).\textquotedblright\ \citep[][pp.\ 37--38]{Walton_1990}.
\end{quote}
The term \textquotedblleft prompt\textquotedblright\ can also be used to describe the function of scientific images: the actual image functions as prompt for a mental image, in the same way a snow fort can function as prompt for an imagined military fort. 
\par The imaginings are prompted by the actual image or fort, following \textquotedblleft principles of generation.\textquotedblright\ These principles guide the mental process of imagination, and form the connection between actual image and mental image: they allow the actual image to \textquotedblleft ignite the spark\textquotedblright\ of the image-ination. \citet{Meynell_2017} gives the following description:
\begin{quote}
    \textquotedblleft [Principles of generation] are the cognitive capacities, habits of mind, and background beliefs that are required, in concert with the text, to produce the appropriate imaginings. (...) Each of these [principles of generation] is trained through practice and application and though some are explicitly stipulated, many are tacitly developed.\textquotedblright\ \citep[][p.\ 504]{Meynell_2017}.
\end{quote}
Although the cognitive processes that create mental images are to a certain degree unique to each individual, principles of generation are shared among (groups of) people because of which authors can use imagery to communicate imaginings to their audience. Through these principles, readers can not only generate mental images based on the actual image, they can also manipulate them. For instance, they can add motion to the image, in such a way that it becomes some sort of animation: a kind of visual, mental simulation. As \citet{Meynell_2020} puts it: \textquotedblleft We can, in our imaginations, remove and include various features and \textquoteleft see\textquoteright\ the many interrelated effects this will have on the pictured state of affairs.\textquotedblright\ \citep[][p.\ 52]{Meynell_2020}. One might argue that a sentence can evoke mental images in a similar way, and while this might be true, \citet{Meynell_2020} states that images can do this in a significantly more efficient way:
\begin{quote}
    \textquotedblleft [S]entences are actually pretty poor at specifying all of the antecedent conditions that are required for a certain state of affairs to come about (...). Pictures do this better because they show their content all at once (...).\textquotedblright\ \citep[][p.\ 52]{Meynell_2020}.
\end{quote}
That is, because the content is all contained in one image, it is more efficient in showing the relation between its elements, and thus forms a more helpful aid to the image-ination.
\par All images can be said to spark the image-ination to a certain degree. For example, the image in Figure \ref{fig3} also helps the forming of mental images, even though its \textquotedblleft main\textquotedblright\ function is best described as a model. After all, for Meynell scientific imagery plays a representational role not dissimilar to Downes' account. However, I want to stress Meynell's account of imagery helping the mental visualization of its audience. In general, as the topic of an article gets more complex, the need for an image that helps the audience mentally visualize the content increases. \citet{Meynell_2017} remarks something similar:
\begin{quote}
    \textquotedblleft When these events are sufficiently complex, surprising or unfamiliar, a picture offers our actual eyes something to work with to help us imagine (...).\textquotedblright\ \citep[][p.\ 509]{Meynell_2017}.
\end{quote}
That is, an image can combine complex elements, providing insight into the \textquotedblleft events\textquotedblright\ being described. When readers see this image, they can form a corresponding mental image, effectively storing the image in their image-ination. Additionally, after the image-ination is triggered by the actual image, the reader can manipulate their mental image (e.g., add motion), which is where the principles of generation come into play. As Meynell states: imagery can give the image-ination of the reader \textquotedblleft something to work with.\textquotedblright\ For some scientific images it appears as if the \textit{main} purpose of the image is to fulfill this function.
\par Interestingly, \citet{Meynell_2018} examines the example of Feynman diagrams, and in particular instances where they do not appear to function as a \textquotedblleft calculational device\textquotedblright\ \citep[see also][]{Meynell_2008}. That is, in these instances they are not used to formulate equations in order to perform calculations regarding the cross-sections of particle interactions, which is their usual purpose \citep[e.g.,][]{Griffiths_2008}. She argues that such examples show that their function as aid to the image-ination is significant, since in these cases this seems to be their only function: 
\begin{quote}
    \textquotedblleft The use of [Feynman diagrams] in contexts where they lack any calculational application (...) is particularly suggestive, implying that the representational role did important epistemic work. Here, the idea that the diagrams were only props for the imagination seems particularly plausible.\textquotedblright\ \citep[][p.\ 463]{Meynell_2018}.
\end{quote}
Again, Meynell uses Waltonian terms and calls these Feynman diagrams \textquotedblleft props for the imagination.\textquotedblright\ After all, it is difficult to argue that these diagrams are propositional: they do not function as \textquotedblleft evidence\textquotedblright\ (in Perini's account) and are not valuable because of their \textquotedblleft truth\textquotedblright. Moreover, although they could be described as model descriptions (in Downes' account) when used to calculate cross-sections, this does not explain their function when they are not used as calculational devices. Meynell's account, however, seems to provide a plausible explanation for their use: the diagrams can be used to help readers form mental images of the described processes. These images do not have to accurately represent the physical particle interactions (which are governed by quantum physics), but can still provide useful insight (i.e., \textquotedblleft something to work with\textquotedblright, as will be elaborated on in Section \ref{sec3.2}).
\subsection{The Stellar Graveyard}
\label{sec2.4}
Can the accounts of Perini, Downes, and Meynell provide an explanation for the Stellar Graveyard plot (Figure \ref{fig1})? First of all, I do not believe that the function of a scientific image can always be decisively categorized as belonging to one of these account. In fact, I think it is conceivable that an image might have multiple functions or one that is not accurately described by any of these three philosophical accounts of scientific imagery. Nevertheless, the accounts of Perini, Downes, and Meynell do shed light on various aspects of the Stellar Graveyard image, which I will investigate below.
\par Firstly, does the Stellar Graveyard plot function as a propositional image? It is true that the figure mainly supports a claim (i.e., a proposition/statement that has a truth-value), namely the claim that the masses of the observed binary black holes are as indicated by the circles in the image. In this regard, the figure itself could be said to have a truth-value. After all, if one of the circles is translated upwards by $20M_{\odot}$, the figure would be \textquotedblleft false\textquotedblright\ since that particular black hole does not have that observed mass. However, if the main function of the Stellar Graveyard plot is to display the compact object masses, then one would expect the figure to be used in scientific arguments similar to the mass distribution shown in Figure \ref{fig2}. But, as has already been mentioned above, the figure is designed sub-optimally for this purpose: any potential structure in the mass distribution cannot be clearly inferred from the individually placed dots. This is why, when the Stellar Graveyard is used in an article, it is usually referred to as a mere display of the observed black hole masses. For example, as stated above, \citet{Mandel_2022} refer to the Stellar Graveyard plot by stating: \textquotedblleft The masses of the black-hole binaries observed to date are shown in Fig. 1, along with masses inferred from X-ray binary observations. Specific system properties are discussed below.\textquotedblright\ \citep[][p.\ 3]{Mandel_2022}. This is typical of the usage of the Stellar Graveyard: it is used to show the observations, but for analyzing the \textquotedblleft specific system properties\textquotedblright\ additional discussion is required. \citet{Broadhurst_2022} show one of the few attempts to use the Stellar Graveyard in their scientific analysis. They do this by modifying the Stellar Graveyard plot and sorting the black hole mergers by their mass to compare the resulting curve with their simulation. However, in showing their final conclusion, they return to an image of a histogram, similar to Figure \ref{fig2}. In general, the Stellar Graveyard, as it is used in articles, does not play an important role in the scientific argument. One might argue that the added value of the Stellar Graveyard plot is simply the fact that it displays the individual black hole masses, as opposed to the total mass distribution shown in a histogram-like image. However, there are multiple ways to display the individual masses that are significantly more suitable for scientific analysis (e.g., Figure 1 of \citeauthor{Disberg_2023} \citeyear{Disberg_2023} or Figure 2 of \citeauthor{LIGO_2023} \citeyear{LIGO_2023}). Because the claim made by the Stellar Graveyard can be displayed in ways that show significantly greater support (cf.\ \textquotedblleft evidence\textquotedblright) for the scientific argument, and these ways are not unknown to astrophysicists, I argue the figure does not fit Perini's description of propositional imagery well. 
\par If the Stellar Graveyard plot is not a propositional image, does it then function as a (description of a) model? There are aspects of the Stellar Graveyard plot that may indicate a model-like function. For instance, the masses of the compact objects are not observed directly but inferred from the shape and intensity of the gravitational wave. In order to do this one needs to model the generation of the gravitational wave by the binary, making several assumptions and idealizations. However, the Stellar Graveyard plot does not display elements of this mathematical model in the sense that for example Figure \ref{fig3} does, and it is certainly not used as a model description. Nevertheless, I argue that there are elements in the figure that appear to be meant to represent the (binary) black holes in this model: (1) the background is black, similar to empty space, (2) the black holes are represented as individual (solid) circles, similar to real (spherical) black holes, and (3) they are ordered somewhat arbitrarily, which could be seen as suggesting a spatial dimension, almost as if it were a \textquotedblleft group picture\textquotedblright\ of the black holes (where the picture might convey a message such as \textquotedblleft We have observed several black holes, and here they are!\textquotedblright). Hence, the Stellar Graveyard plot seems to aim for a similarity with the actual black holes, in some ways comparable to a similarity relation between model and object. However, it is difficult to argue that the main purpose of the Stellar Graveyard figure is to display accurate depictions of black holes and neutron stars. The virtue of the image does not lie in the accuracy of its representation, in fact the dots are very low-level representations of neutron stars and black holes that are not referred to in texts. Downes stresses how scientific images can display models whose accuracy (i.e., similarity to the objects they describe) is what makes science successful, but---in contrast to for example the black hole image produced by \citet{EHT_2022}\footnote{See also their website: \href{https://eventhorizontelescope.org}{eventhorizontelescope.org}.}---this is not the main function of the Stellar Graveyard image.
\par I argue that Meynell's account of scientific imagery can give a much more accurate description of the Stellar Graveyard's function. That is, I believe that even though this image is not very suitable for scientific analysis, it is more effective in appealing to the image-ination than more technical images such as Figure \ref{fig2}. The fact that the black holes are pictured as circles, together with an arrow which shows their merger, makes it easier for the reader to picture the moving black holes and their merger which results in a more massive black hole. I therefore argue that the (somewhat trivial) similarity between the image and actual black holes, as mentioned above, does not suffice to consider the image itself an astrophysical model (or a model description of another astrophysical model) that establishes an accurate representational relationship with a certain object in the world (i.e., the black holes) but instead makes the figure an effective prompt for the image-ination. The accuracy of the representation is less important than in Downes' account. Similarly, in the example of Walton (mentioned above) it is not that the value of the snow fort lies in its accurate resemblance to a real fort, but in its appeal to the image-ination of the children (through principles of generation). In general, the argument I make regarding the Stellar Graveyard is similar to the example of Feynman diagrams given by Meynell: the fact that the image does not function as a \textquotedblleft calculational device\textquotedblright\ or support a scientific conclusion in any other shape or form, makes the idea that its value lies outside of the scientific argument (i.e., in its appeal to the image-ination of the reader) more plausible. This does not play a role in the scientific analysis but instead describes how scientists take into account how their images affect their audience, and is not well-described by the accounts of Perini and Downes (who focus on the scientific argument/analysis).
\newpage
\section{Imagination and Astrophysics}
\label{sec3}
Since the function of the Stellar Graveyard is well-described as an aide to the image-ination, I will describe the scientific method of astrophysics and the role the image-ination can play in astrophysical research (Section \ref{sec3.1}). I will also turn to the philosophical theories of \citet{DeRegt_2017} and \citet{Meynell_2020} regarding scientific understanding to argue that imaginative imagery is linked to scientific understanding (Section \ref{sec3.2}). Then, I will give several more examples of astrophysical images that appear to exhibit this function (Section \ref{sec3.3}). 
\subsection{The Cosmic Laboratory}
\label{sec3.1}
Within the philosophy of science, some areas of physics have been of particular interest: many articles and books have been written on philosophical issues in, for example, quantum mechanics or relativity. However, the philosophical exploration of the methods used in astrophysics has been somewhat limited, compared to these other disciplines \citep{Anderl_2016}. Recently, however, the first book on the philosophy of astrophysics has been published \citep{Boyd_2023}, which is of course a step in the right direction. Moreover, astrophysics plays a unique role in the work of \citet{Hacking_1983}, who argues for anti-realism with regard to the objects studied by astrophysicists, because scientists---generally speaking---cannot manipulate astrophysical objects in experiments. The astrophysical method, according to \citet{Hacking_1989}, is special because it cannot rely on experiments, and therefore has to resort to \textquotedblleft saving the phenomena.\textquotedblright\ This term dates back to the ancient Greeks, and describes how observations that contradict established theory can be \textquotedblleft saved\textquotedblright\ by adjusting the used model. Hacking gives the following example: if ancient astronomers observe planets moving in a way that contradicts their geocentric model, they can attempt to \textquotedblleft save\textquotedblright\ this by adding more epicycles to their model. This \textquotedblleft phenomena-saving\textquotedblright\ is a characteristic aspect of astrophysics:
\begin{quote}
    \textquotedblleft Hence saving the phenomena seems an entirely subsidiary aspect of scientific activity. There is one, and perhaps only one, branch of science where the tag \textquoteleft to save the phenomena\textquoteright\ has a central place: astronomy and astrophysics.\textquotedblright\ \citep[][p.\ 577]{Hacking_1989}.
\end{quote}
Astrophysics has a unique dependence on observation, and this affects the range of research methods available to the astrophysicist. Because of this, phenomena-saving has a \textquotedblleft central place\textquotedblright\ in astrophysics. \citet{Hacking_1989} argues that, although the methods of most natural sciences have historically undergone significant changes due to technological progress, the astrophysical method can still be described as phenomena-saving:
\begin{quote}
    \textquotedblleft The technology of astronomy and astrophysics has changed radically since ancient times, but \textit{its method remains exactly the same}. Observe the heavenly bodies. Construct models of the (macro)cosmos. Try to bring observations and models into line.\textquotedblright\ \citep[][p.\ 577, italics in original]{Hacking_1989}.
\end{quote}
This aligns well with the description of \citet{Downes_2011}: astrophysics is successful to the extent that astrophysical models are similar to the (observed) cosmos.
\par \citet[][p.\ 820]{Anderl_2018} also states that this method is characteristic for astrophysics. This is, for instance, reflected by the fact that within academia there is a relatively clear distinction between observational and theoretical astrophysics. The interplay between model and observation entails what \citet{Anderl_2016} calls a \textquotedblleft Sherlock Holmes\textquotedblright\ methodology.\footnote{Interestingly, \citet{Einstein_1938} refer to the author of the Sherlock Holmes novels when comparing science to \textquotedblleft a mystery story.\textquotedblright\vspace{-5mm}} She points out that astrophysicists only have access to the current state of the universe, and therefore have to reconstruct a causal history in order to explain this state \citep[see also][]{Anderl_2021}. For example, stellar evolution typically occurs on timescales too long to observe directly, but by observing a large number of stars evolutionary patterns can be extracted \citep[e.g.][]{Gaia_2018}. Because of this tracing back of clues, astrophysics has been called a \textquotedblleft historical science\textquotedblright\ as opposed to an \textquotedblleft experimental science\textquotedblright\ \citep{Cleland_2002}. The \textquotedblleft historical\textquotedblright\ analysis of astrophysical phenomena is possible because we can observe many of these phenomena in different stages of their evolution, in the \textquotedblleft Cosmic Laboratory\textquotedblright\ \citep[][p.\ 2]{Anderl_2016}. Astrophysicists therefore depend on the \textquotedblleft Cosmic Laboratory\textquotedblright\ to present them phenomena that harbor clues about their origin.
\begin{figure}
     \centering
     \begin{subfigure}[b]{0.38\textwidth}
         \centering
         \includegraphics[width=\textwidth]{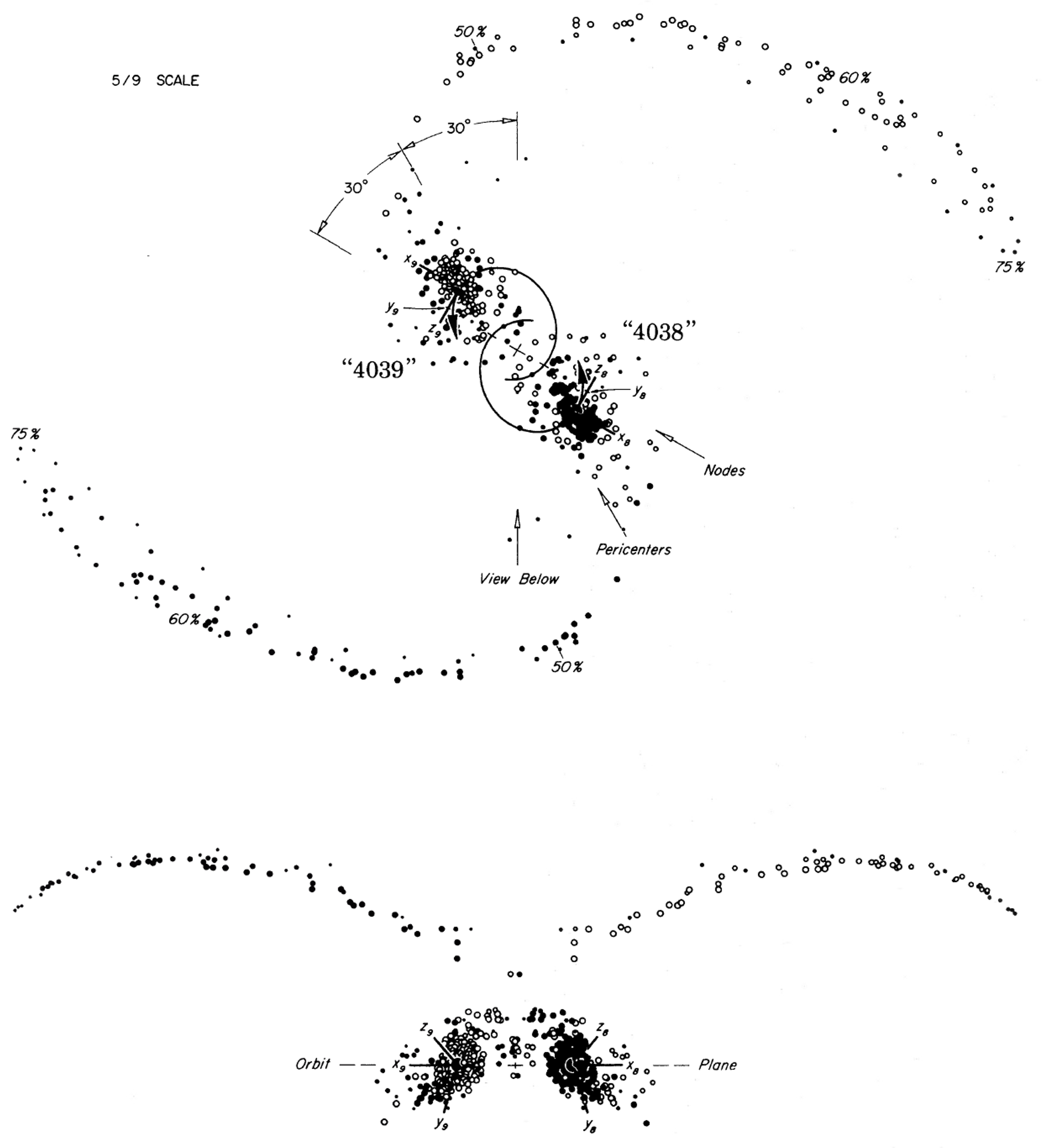}
         \caption{Toomre simulation}
         \label{fig4a}
     \end{subfigure}
     \hfill
     \begin{subfigure}[b]{0.61\textwidth}
         \centering
         \includegraphics[width=\textwidth]{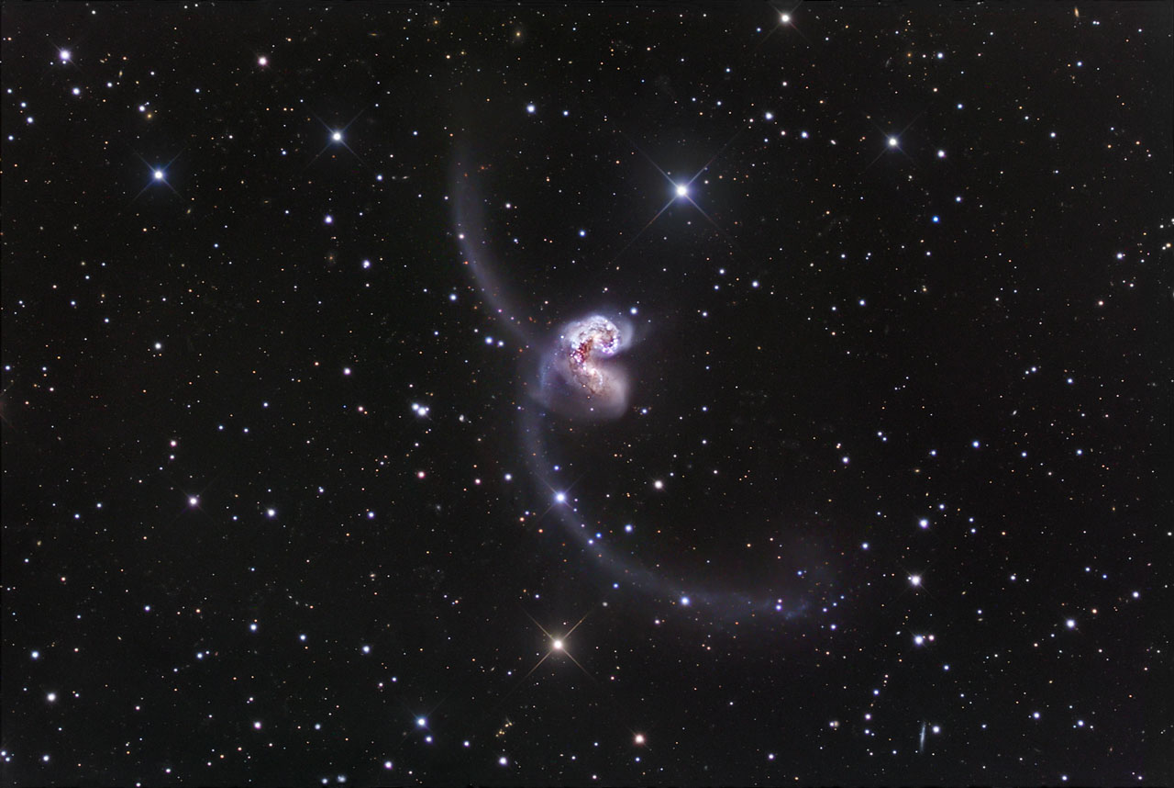}
         \caption{Antennae galaxies}
         \label{fig4b}
    \end{subfigure}
    \caption{Part (a): simulation by \citet[][i.e., their Fig.\ 23, \textcopyright\,AAS---reproduced with permission]{Toomre_1972}, which was designed to resemble the Antennae galaxies \citep[cf.][]{Lahen_2018}. The panel shows both a top view and a side view of the two galaxies. Part (b): an image of the actual observed Antennae galaxies, taken with the Hubble Space Telescope (credit: \href{https://esahubble.org/images/heic0812c/}{Robert Gendler}).}
     \label{fig4}
\end{figure}
\par An example of aligning model and observation in astrophysics, which shows the \textquotedblleft Sherlock Holmes\textquotedblright\ method, is the simulation by \citet{Toomre_1972}. They simulate colliding galaxies by calculating the gravitational attraction of the stars to the galactic nuclei while neglecting the gravitational attraction between the stars, which reduces the computational power needed for the simulation. This method, as it turns out, is successful in recreating galaxies with irregular shapes, such as tail-like structures. In order to do this, there are many parameters which can be varied in their simulation, such as the orientation of the galactic planes and the velocities of the interacting galaxies. Figure \ref{fig4a} shows the results for one of their simulations, which attempts to reproduce shapes similar to the Antennae galaxies (shown in Figure \ref{fig4b}). The tails of the Antennae galaxies appear to cross, which could not be explained by the established theory of galaxy formation. \citet{Toomre_1972} therefore expected finding an explanation for this phenomenon to be problematic. However, they state that \textquotedblleft Figure [\ref{fig4a}] says we need not have worried.\textquotedblright\ \citep[][p.\ 659]{Toomre_1972}. They implicitly argue that because in their simulation the tails do not cross in the third dimension, this is the probable spatial configuration of the actual Antennae galaxies. Hence, these galaxies do not challenge the theory of galaxy formation after all: they are \textquotedblleft saved\textquotedblright\ by the Toomre simulation. \citet{Schemmel_2023} discusses the Toomre simulation and the function of, for instance, Figure \ref{fig4a} in their argument. He concludes: \textquotedblleft Images have thus provided a pivotal link between empirical appearance and theoretical construction\textquotedblright\ \citep[][p.\ 315]{Schemmel_2023}, stressing how the image reconciles observation and theory, effectively saving the phenomenon.
\par In addition to actual images (such as Figure \ref{fig4}), mental images can also provide this \textquotedblleft pivotal link\textquotedblright\ between model and observation. That is, Sherlock Holmes might use abduction to infer the most likely scenario of a certain crime from a set of clues (e.g., hoof prints or mud stains), where the image-ination can be a useful tool to go over possible theoretical scenarios and link them to the empirical evidence. Similarly, when observational astronomers are faced with transient signals (i.e., signals that appear and disappear on a relatively short timescale), for example, the image-ination can be a useful tool in trying to come up with an explanation for these observations.\footnote{\textquotedblleft The great part of astronomy is the imagination that’s been necessary to guess what kind of structures, what kinds of things, could be happening to produce the light and the effects of the light, and so on, of the stars we do see.\textquotedblright\ \citep[][p.\ 86]{Feynman_2015}.\vspace{-5mm}} In such a case, the \textquotedblleft clues\textquotedblright\ often entail the intensities of the observed light, per frequency, and how these evolve over time. Although this information is relatively limited, astrophysicists have linked transient signals to a wide variety of cosmic events, such as stars being torn apart by black holes \citep[\textquotedblleft tidal disruption events,\textquotedblright][]{Frank_1976}, the merger of two neutron stars \citep[\textquotedblleft short-duration gamma-ray bursts,\textquotedblright][]{Eichler_1989}, jets caused by the collapse of a massive, metal-poor star \citep[\textquotedblleft long-duration gamma-ray bursts,\textquotedblright][]{Woosley_1993}, or pulsational mass-ejections in supernova explosions \citep[\textquotedblleft pulsational pair-instability supernovae,\textquotedblright][]{Woosley_2007}, although there are many transients of which the origin is still highly uncertain \citep[e.g.\ \textquotedblleft fast X-ray transients,\textquotedblright][]{Connors_1986}. I argue that the image-ination can be particularly useful in astrophysics, since---due to its method of phenomena-saving---it has a constant search for new or adjusted models that can explain observations. Mental imagery is a helpful tool for this, since it presents its content all at once (as discussed in Section \ref{sec2.3}) and can provide insight in how the spatial structure of the system has shaped the observed signal. Imagery that sparks the image-ination can therefore be used to communicate this kind of insight. Most astrophysical objects (e.g.\ stars, nebulae, and galaxies) are macroscopic, and it makes sense to talk about their visual appearance (as in Figure \ref{fig4}). This sets astrophysics apart from more abstract disciplines, such as quantum physics or mathematical physics, where spatial structures and visualization play a less prominent role. 
\subsection{Understanding through Imagination}
\label{sec3.2}
Richard Feynman often commented on the relationship between science and imagination. For example, he notes that, in contrast to how imagination may be used in everyday life, in science the purpose of the imagination is to \textquotedblleft comprehend\textquotedblright\ the events that are, for instance, present in the Cosmic Laboratory:
\begin{quote}
    \textquotedblleft Our imagination is stretched to the utmost, not as in fiction, to imagine things which are not really there, but just to comprehend those things which \textit{are} there.\textquotedblright\ \citep[][pp.\ 127--128, italics in original]{Feynman_1965}.
\end{quote}
In other words, for Feynman achieving understanding of the outside world is a goal of the scientific imagination, and this means that the latter is in a sense constrained to match the outside world, to a certain degree.\footnote{In his biography of Feynman, author James \citet{Gleick_1992} describes this as follows: \textquotedblleft The kind of imagination that takes blank paper, blank staves, or a blank canvas and fills it with something wholly new, wholly free---that, Feynman contended, was not the scientist’s imagination. (...) For Feynman, the essence of the scientific imagination was a powerful and almost painful rule. What scientists create must match reality. It must match what is already known. Scientific creativity, he said, is imagination in a straitjacket.\textquotedblright\ \citep[][pp.\ 509--510]{Gleick_1992}.\vspace{-5mm}} Michael \citet{Stuart_2017} also describes the connection between imagination and understanding, and notes how this can be achieved through an \textquotedblleft exercise of the imagination\textquotedblright:
\begin{quote}
    \textquotedblleft [W]here novel understanding is produced, it is often due to creating a connection between some theoretical structure(s) of science and existing knowledge, skills or experience, via an exercise of the imagination.\textquotedblright\ \citep[][p.\ 27]{Stuart_2017}.
\end{quote}
Although it goes beyond the scope of this article to give a comprehensive overview of the debate on scientific understanding and its connection to imagination, I will illustrate how imagery and image-ination might be used to further understanding, using the work of \citet{DeRegt_2017} and \citet{Meynell_2020}.
\par Henk de Regt formulates a pragmatic account of scientific understanding, in which a phenomenon is understood scientifically if there is an adequate explanation for it, and this explanation is based on an \textquotedblleft intelligible\textquotedblright\ theory \citep[][p.\ 92]{DeRegt_2017}. De Regt argues that understanding a scientific theory means being able to \textquotedblleft use\textquotedblright\ it \citep[][p.\ 91]{DeRegt_2017}, and he provides a possible criterion for this kind of intelligibility:
\begin{quote}
    \textquotedblleft A scientific theory T (in one or more of its representations) is intelligible for scientists (in context C) if they can recognize qualitatively characteristic consequences of T without performing exact calculations.\textquotedblright\ \citep[][p.\ 102]{DeRegt_2017}.
\end{quote}
This is an important notion: a scientific theory is intelligible if scientists can grasp the \textquotedblleft qualitatively characteristic consequences\textquotedblright\ of the theory \citep[see also][]{DeRegt_2005}. These characteristic consequences are not deduced through \textquotedblleft exact calculations,\textquotedblright\ but instead scientists need to have an idea of the general implications of the theory. I believe the image-ination plays a crucial role in the process of determining the characteristic consequences of a theory: scientists can \textquotedblleft mentally simulate\textquotedblright\ how a situation will evolve, based on the scientific theory. \citet{Murphy_2022} gives a similar description:
\begin{quote}
    \textquotedblleft For example, for De Regt, part of the intelligibility of a theory comes from getting a feeling of the consequence of the theory in a concrete situation (...). And this can involve \textquoteleft seeing\textquoteright\ these consequences, that is, imagining them in a visual way.\textquotedblright\ \citep[][p.\ 6]{Murphy_2022}. 
\end{quote}
In this way, scientists can use their imagination to assess the intelligibility of a theory: if they can visually imagine the characteristic consequences, the theory is intelligible. Adding images that \textquotedblleft spark\textquotedblright\ the imagination can, therefore, help the reader assess the intelligibility of the theoretical statements in an article.
\par De Regt's notion of intelligibility is somewhat black and white: scientists either \textit{can} or \textit{cannot} recognize the characteristic consequences of a theory, which does not allow for speaking about \textquotedblleft degrees of intelligibility.\textquotedblright\ Because of this, his definition makes it difficult to analyze whether using one's image-ination can increase understanding. After all, scientists can perhaps use their image-ination to increase the ways in which they can use the theory, but an increased ability does not lead to an increase in intelligibility, according to De Regt's definition. This is why I would argue that the definition of \citet{DeRegt_2017} should be altered in such a way that the intelligibility of a theory is linked to \textquotedblleft the extent to which\textquotedblright\ a scientist can recognize the qualitatively characteristic consequences of a theory. Regardless, I argue that it is, in fact, possible for the image-ination to further scientific understanding. After all, in a more general sense, forming mental images of a theoretical statement can reveal connections to already established knowledge, possibly increasing the number of ways scientists can use the theory---which could be called a \textquotedblleft search for understanding\textquotedblright\ \citep[][p.\ 22]{Stuart_2017}---and therefore increases this Regtian sense of understanding.
\par \citet{Meynell_2020} also views understanding as deeply connected to imagery and image-ination. For her, images are useful because they show all of their content at once (as discussed in Section \ref{sec2.3}), which reveals spatial structure and (implicit) causal relationships. These are essential for understanding, because of which Meynell argues that the \textquotedblleft characteristic content\textquotedblright\ of understanding is pictorial in nature, similarly to how the characteristic content of knowledge is propositional \citep[][p.\ 40]{Meynell_2020}. Understanding, then, is the ability to manipulate these structures and causal relations (in a \textquotedblleft plausibility space\textquotedblright) and \textquotedblleft seeing\textquotedblright\ how these modifications affect the situation. Meynell describes this as follows:
\begin{quote}
    \textquotedblleft To understand something is to be able to see the component parts in relation to each other and as they relate to the whole. (...) If the object of one’s understanding is a state of affairs or phenomenon, as is typical in the sciences, causal relations are put into this kind of relational web. Those who deeply understand causal phenomena can see a kind of plausibility space where they can consider how things might have been different and appreciate, for any given change, what would follow from it with a view to the multiple relations that might be modified by the change.\textquotedblright\ \citep[][p.\ 58]{Meynell_2020}.
\end{quote}
In other words: to really understand something, according to Meynell, means to be able to consider \textquotedblleft what-if-things-had-been-different scenarios\textquotedblright\ \citep[][p.\ 52]{Meynell_2020} and their effects on the situation. These effects can be compared to De Regt's notion of qualitatively characteristic consequences: they are not established through explicit calculations, but rather through some kind of qualitative mental simulation (cf.\ Stuart's \textquotedblleft exercise of the imagination\textquotedblright).
\par Meynell's description of going over different scenarios and image-ining their qualitative consequences resembles the astrophysicist's search for models that can save their observations, as described in Section \ref{sec3.1}. Truly understanding the relationship between model and observation, in this account of scientific understanding, is closely related to the astrophysicist's ability to \textquotedblleft see\textquotedblright\ how the qualitative characteristic consequences of the theoretical model shape the (predicted) observation and how changing the model would alter it. Although Meynell refers to these changes as \textquotedblleft non-factive\textquotedblright\ \citep*{Meynell_2018} or \textquotedblleft counterfactual\textquotedblright\ \citep*{Meynell_2020}, I would prefer to consider them \textquotedblleft hypothetical\textquotedblright\ in this context, since a fiducial model (i.e., a model that variations are compared to) is not necessarily \textquotedblleft factual.\textquotedblright\ Moreover, when writing a paper scientists aim to present their argument in such a way that it is intelligible to readers (i.e., other scientists). After all, when astrophysicists publish articles, they are doing more than presenting their scientific work. For instance, they also want to convince the scientific community that their work is correct and an accurate description of reality, because of which they are motivated to communicate their message as clearly as possible, in order to avoid unnecessary skepticism due to a misunderstanding. Authors can aid the understanding of the reader by helping them form mental images of the scientific topic through images included in the article, because of which this imagery becomes an important tool for scientific understanding in general \citep[cf.][p.\ 256]{DeRegt_2017}. Moreover, the thought processes that motivated scientists to pursue a certain research project are relevant for the scientific article, since they may reveal how intuitively (e.g., based on mental images) the scientists already expect certain conclusions. This, of course, is not a scientific justification for these conclusions. However, if scientists can communicate their intuitive mental images to readers, the latter may form their own mental images and is then perhaps more easily persuaded by the scientific argument \citep[e.g., Figures 8 and 9 from][]{Disberg_2024a}.
\subsection{Astrophysical Imagery}
\label{sec3.3}
I have argued that the function of the Stellar Graveyard plot cannot be satisfactorily described as either propositional or a model (description). Instead, I have posed that the main function of this image is to evoke mental imagery of the binary black holes. The Stellar Graveyard plot is a useful example to illustrate this function, but there are many more images present in astrophysical articles that also appear to be well-described by Meynell's account. I will therefore discuss three additional examples of this kind of imaginative imagery, concerning topics related to the Stellar Graveyard image.
\par
\begin{figure}
     \centering
     \begin{subfigure}[b]{0.47\textwidth}
         \centering
         \includegraphics[width=\textwidth]{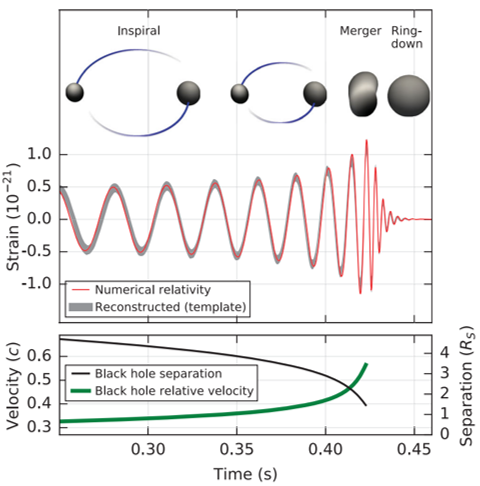}
         \caption{}
         \label{fig5a}
     \end{subfigure}
     \hfill
     \begin{subfigure}[b]{0.52\textwidth}
         \centering
         \includegraphics[width=\textwidth]{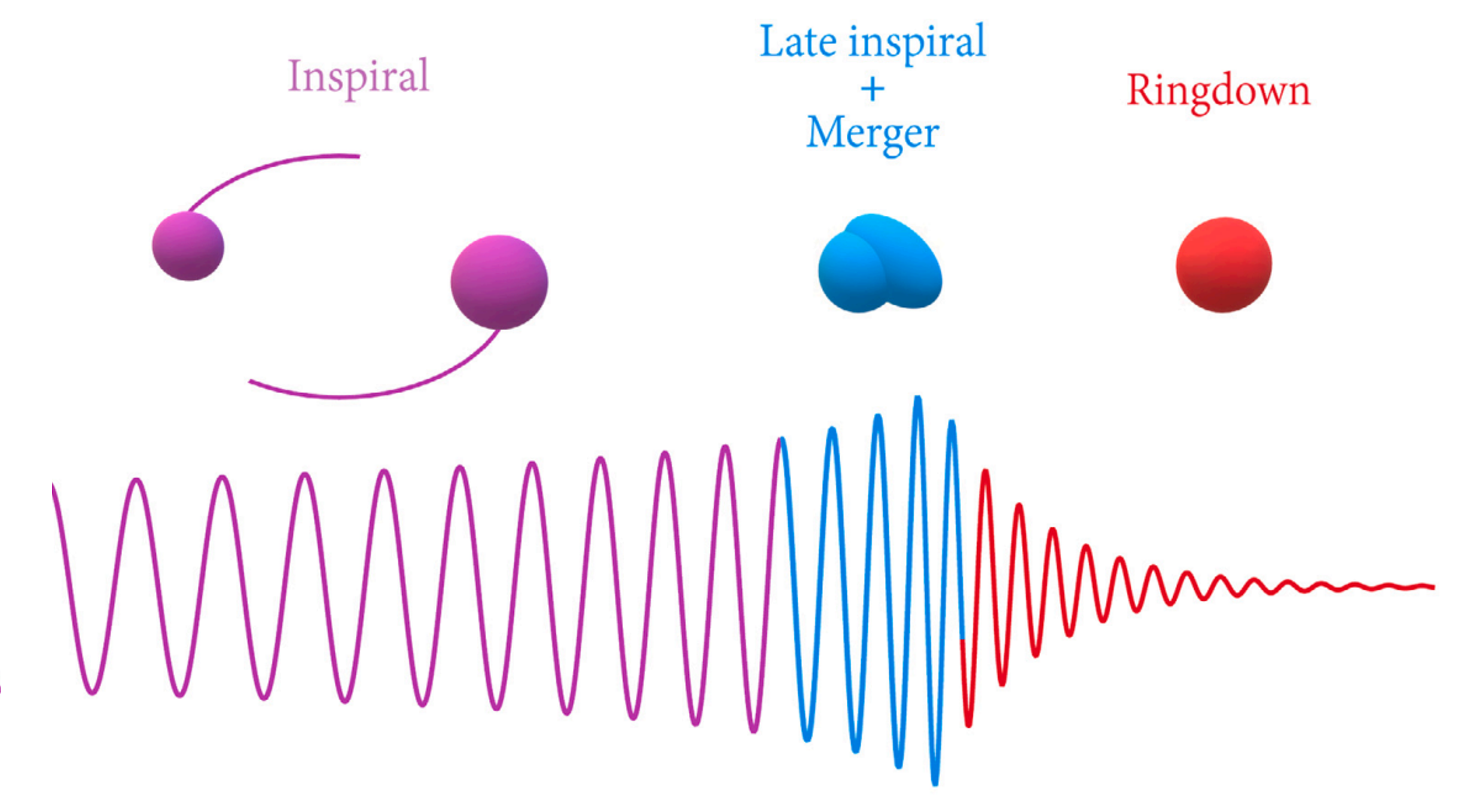}
         \caption{}
         \label{fig5b}
    \end{subfigure}
    \caption{First observation of a gravitational wave resulting from a black hole merger. Part (a) shows a reconstruction of the signal and predictions made using general relativity \citep[Fig.\ 2 of][reproduced following the CC BY 3.0 license]{LIGO_2016}, along with diagrams of the simulated black hole horizons. Part (b) shows a theoretical gravitational wave signal accompanied by diagrams of the merger \citep[amended version of Fig.\ 1 of][reproduced following the CC BY 4.0 license]{Cattorini_2024}.}
     \label{fig5}
\end{figure}

 The first example consists of images displaying gravitational waves detected from black hole mergers, which are waves in the fabric of spacetime, caused by the spiraling in and eventual merging of two massive objects (e.g., black holes). As shown in Figure \ref{fig5}, they are sometimes accompanied by images of the merging black holes. Figure \ref{fig5a}, for instance, is part of the publication of the first observed gravitational wave from a binary black hole merger \citep{LIGO_2016} and shows a comparison between a numerical simulation of a merging black hole binary and the observation. The top part shows how their simulated black holes spiral inwards and merge into a more massive black hole, whereas the bottom panel shows the evolution of the distance between the black holes and their relative velocity. One might argue that the black hole images resulting from the numerical simulation are model descriptions, since they are meant to be accurate representations of the actual black holes that produces the gravitational wave signal. However, the authors refer to Figure \ref{fig5a} as follows: \textquotedblleft [The figure] shows the calculated waveform using the resulting source parameters.\textquotedblright\ \citep[][p.\ 3]{LIGO_2016}; they only refer to the gravitational wave and do not even mention the black hole images in their main text. This is why I do not believe that these images are merely valuable due to the simulation being an accurate representation of the actual black holes.
\par Instead, I argue that Meynell's account of imagery and image-ination provides a significantly better description for the function of these diagrams. They function as a Waltonian \textquotedblleft prompt\textquotedblright\ for the generation of mental images of the spiraling black holes. While these image-ined black holes spiral inward, they emit gravitational waves and therefore lose energy, resulting in a decrease in the distance between the black holes and an increase in their velocities. That is, the qualitatively characteristic consequences of the gravitational wave emission are easily imagined through the black hole imagery, effectively reconciling model and observation. For example, in their review paper \citet{Bailes_2021} display a version of Figure \ref{fig5a} that also includes these black hole images and refer to them as: \textquotedblleft (...) a simulation of the merger produced using numerical relativity to illustrate the evolution of the black hole event horizons as the system coalesces and merges.\textquotedblright\ \citep[][p.\ 346]{Bailes_2021}. Here, I argue that to \textquotedblleft illustrate\textquotedblright\ the merger, the fact that the size and shape of the horizons are simulated to be accurate representations of the actual black holes is not the most important feature of these images. Instead, the main function of these images is that they are useful tools for showing the spatial structure of the system and the causal relations between the gravitational waves and the distances between the black holes and their velocities (which aligns well with Meynell's account). For example, in their review paper \citet{Cattorini_2024} use similar images accompanying a theoretical gravitational wave \citep[see also, e.g.,][]{Baumgarte_2011,Miller_2016,Salcido_2016,Isoyama_2021,Santos_2022}. These images are not accurate simulations but instead qualitative diagrams to illustrate the black hole merger (as shown in Figure \ref{fig5b}). They use these images to list the stages of the merger event and provide insight into how they qualitatively shape the gravitational wave signal. The images do not lose their utility when replaced by these diagrams: their use is not to provide accurate representations but to function as an aide for the image-ination.
\par
\begin{figure}[!ht]
    \centering 
    \includegraphics[width=.6\linewidth]{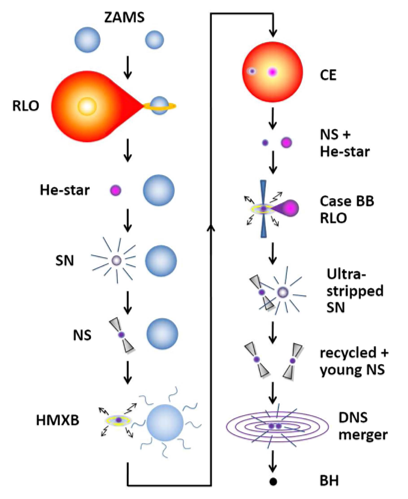}
    \caption{Van den Heuvel diagram of a binary evolution model that described the formation of two neutron stars (NSs) that eventually merge into a black hole (BH) while producing gravitational waves \citep[Fig.\ 1 of][\textcopyright\,AAS---reproduced with permission]{Tauris_2017}. The model includes different episodes of mass transfer between the two stars (RLO), supernovae (SNe), and a common envelope phase (CE).}
    \label{fig6}
\end{figure}

 The second example concerns theoretical descriptions of the evolutionary stages that binary stars go through before ending up as, for example, a black hole binary. Articles on this topic often use images that are commonly referred to as \textquotedblleft Van den Heuvel diagrams\textquotedblright\ \citep[e.g.,][]{Flannery_1975,Van_den_Heuvel_1981}, which show these different stages in the binary evolution, depicting episodes of mass transfer and supernovae \citep[see also, e.g.,][]{Vigna-Gomez_2018,Marchant_2019,Mandel_2022,Van_Son_2022,Tauris_2023,Wagg_2025}. Figure \ref{fig6} shows a Van den Heuvel diagram used in the article of \citet{Tauris_2017}: the circles represent stars or stellar remnants (such as black holes), and the diagrams also show how matter can flow from one star to the other, or form a common envelope (engulfing both objects). Which particular evolutionary stages a certain binary goes through depends on the initial properties such as masses, composition, and separation; meaning there are various evolutionary pathways that can be depicted in a Van den Heuvel diagram. Similarly to the images in the Stellar Graveyard and Figure \ref{fig5}, Van den Heuvel diagrams function as illustrations that are usually not explicitly referred to in the respective texts. That is, when these diagrams are used the authors often also discuss the evolutionary steps in their text, and refer to the Van den Heuvel diagram as an illustration of this without linking the components of the image to the relevant concepts. For example, \citet{Tauris_2017} use the Van den Heuvel diagram shown in Figure \ref{fig6} to illustrate a possible formation of a double neutron star (DNS) system, and refer to the image as: \textquotedblleft In [Figure \ref{fig6}], we show an illustration of the formation of a DNS system.\textquotedblright\ \citep[][p.\ 3]{Tauris_2017}, after which they discuss the evolution of the system without referring to the image.
\par Van den Heuvel diagrams are not used in a way similar to, for example, Figure \ref{fig3}, where elements displayed in the image are explicitly used in the scientific analysis. Instead, they function as illustrations of the binary evolution model that is being used to explain, for instance, the properties of a compact binary. These illustrations, I argue, are effective aides for the image-ination, prompting mental imagery of binary stars are the matter that is transported within or expelled from the system. The prompted mental imagery, similarly to the examples mentioned before, is useful for image-ining the qualitatively characteristic consequences of the physics involved, such as the fact that a common envelope will shrink the orbit and bring the objects closer together. Of course, listing the different steps in the text can also highlight this fact, but (mental) images are much more effective in revealing the structure of the entire system (as described by Meynell). For this, the diagrams and the corresponding mental images do not need to be (accurate) representations of the actual binary stars: it is not necessary that, for example, the scale of the pictured system is accurate. Moreover, the mental images of the binary system are efficient tools for considering hypothetical \textquotedblleft what-if-things-had-been-different scenarios\textquotedblright, as described by \citet{Meynell_2020}, such as the hypothetical effects of rapid rotation causing chemical mixing and therefore changing the stellar structure \citep[e.g.,][]{Mandel_2016} or adding a third star \citep[e.g.,][]{Kummer_2025}.
\par
\begin{figure}
     \centering
     \begin{subfigure}[b]{0.5\textwidth}
         \centering
         \includegraphics[width=\textwidth]{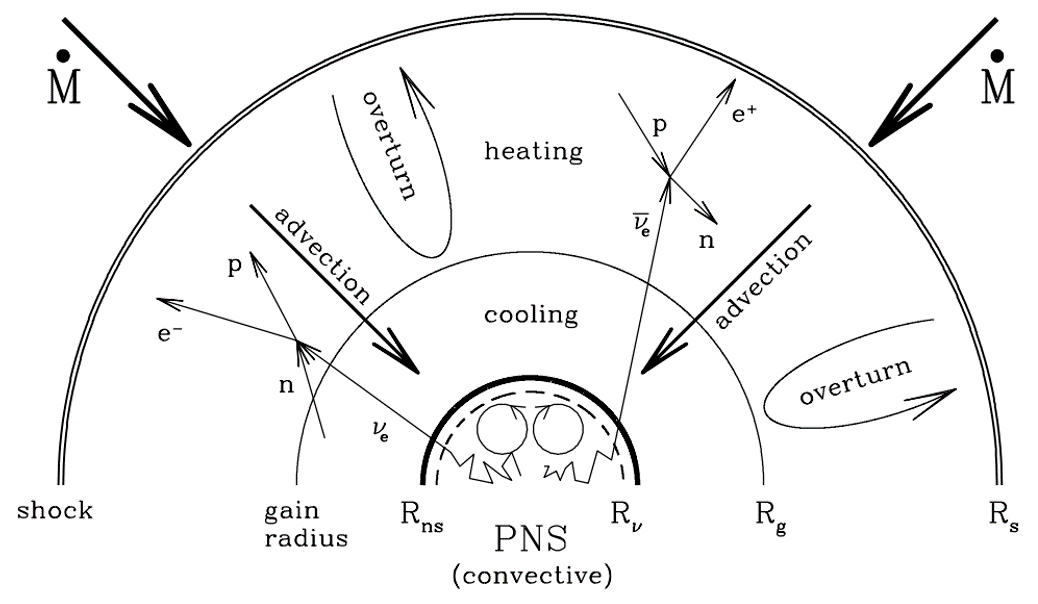}
         \caption{}
         \label{fig7a}
     \end{subfigure}
     \hfill
     \begin{subfigure}[b]{0.49\textwidth}
         \centering
         \includegraphics[width=\textwidth]{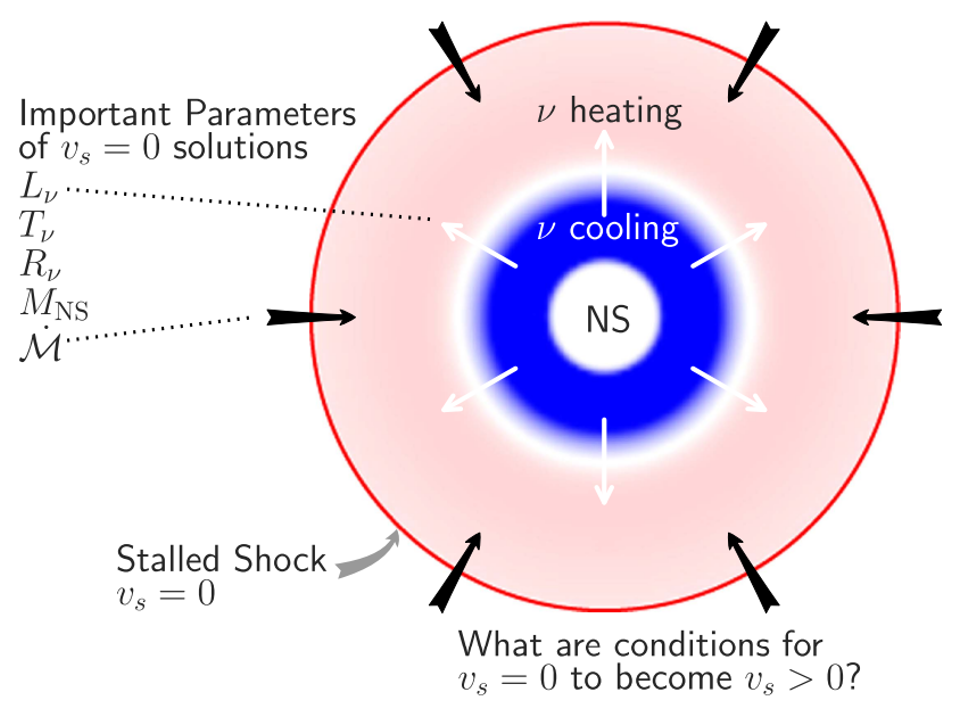}
         \caption{}
         \label{fig7b}
    \end{subfigure}
    \caption{Models of supernova explosions. Part (a) shows the model of \citet[][i.e., their Fig.\ 1---reproduced with permission]{Janka_2001}, which contains a shockwave traveling away from the core as well as particle reactions, and radii relevant for the calculations made in his article. Part (b) shows a similar image from \citet[][i.e., their Fig.\ 1, \textcopyright\,AAS---reproduced with permission]{Murphy_2017}, which has fewer details but lists relevant parameters.}
     \label{fig7}
\end{figure}

 The last example consists of images of supernova explosions that occur at the end of the nuclear burning life of a massive star, as shown in Figure \ref{fig7} (and also depicted in the diagrams of Figure \ref{fig6}). These images show the factors that are being taken into account in describing the structure of the collapsing stellar core and the resulting shockwave. Figure \ref{fig7a}, for example, displays nuclear reactions occurring in the core, as well as several radii that are relevant for determining the evolution of the shockwave. This way, the image has a function similar to Figure \ref{fig3}: it shows a model and defines radii that are explicitly used in the calculations present in the article. Figure \ref{fig7b}, in contrast, shows a similar but more simplistic diagram of a supernova explosion. The authors, \citet{Murphy_2017}, investigate the problem of finding parameters for which the shock velocity is positive, resulting in a successful supernova explosion. They refer to the image as: \textquotedblleft See [Figure \ref{fig7b}] for a schematic of the boundary value problem and the most important parameters of the problem.\textquotedblright\ \citep[][p.\ 3]{Murphy_2017}, after which they discuss the effects of the parameters listed in the figure. Similarly to the previous examples, the authors do not use specific components of the image in their analysis (in contrast to Figure \ref{fig7a}).
\par The argument I make regarding Figure \ref{fig7b} is similar to the arguments regarding the Stellar Graveyard and Figures \ref{fig5} and \ref{fig6}, and to the argument \citet{Meynell_2018} makes regarding certain Feynman diagrams (as discussed in Section \ref{sec2.3}): this image is not explicitly used in the scientific argument being made in the article, which makes it more plausible that it functions as an aide for the image-ination. That is, displaying the spatial structure of the supernova can help the reader mentally visualize the explosion and the shockwave propagating through the dying star, and might provide insight in, for example, the causal relation between the outer layers of the stars falling onto the (proto-)neutron star and the shockwave velocity. This mental imagery can be a tool used by astrophysicists to go over hypothetical scenarios and their qualitatively characteristic consequences, such as a scenario in which the outer (hydrogen) layer of the star is stripped by a binary companion and the effects this might have on the outcome, and these kinds of considerations are relevant for scientific understanding as described by \citet{DeRegt_2017} and \citet{Meynell_2020}. Figure \ref{fig7a}, in contrast, is in fact used in the scientific analysis and its main value does lie in the accuracy with which it represents a supernova; it can therefore be reasonably well described as a model description. However, there are strong similarities between Figures \ref{fig7a} and \ref{fig7b} and I argue that Figure \ref{fig7a} can be an aide to the image-ination in a similar way that Figure \ref{fig7b} can. The two images in Figure \ref{fig7} show how there is no clear divide between model images and imaginative images: an image can function as a model description as well as an imaginative image \citep[see also, e.g., Figure 3 of][]{Disberg_2024b}.
\par The main purpose of the imaginative work aided by these images is connected to understanding. I am of course not claiming that without these images astrophysicists would not be able to form mental images of the discussed concepts and models (which are often relatively simplistic), nor am I claiming that without the prompted mental images they would not be able to understand them. I do argue, however, that the images discussed in this section show that (1) visual thinking / the image-ination is prevalent in astrophysics, (2) astrophysicists use images in their articles to aid this, (3) the mental images are meant to further astrophysical understanding, and (4) this imagery and its connection to the image-ination and understanding is well-described by Meynell's work.
\newpage
\section{Conclusion}
\label{sec4}
In this article, I have investigated the relationship between imagination, understanding, and imagery, within the context of astrophysical literature. In particular, the preceding sections have established the following:
\begin{itemize}
    \item [\ref{sec1}.] The Stellar Graveyard plot (Figure \ref{fig1}) is commonly used to represent the binary black holes observed through gravitational waves \citep[e.g.,][]{Mandel_2022}. This figure, however, is a sub-optimal way of displaying the masses of these black holes for scientific analysis, which means that there must be something else making this image valuable.
    \item [\ref{sec2}.] In general, the images used in scientific articles can have a variety of functions. They can, for instance, function as a propositional claim which can be seen as \textquotedblleft evidence\textquotedblright\ supporting the scientific conclusion of the article \citep[][Section \ref{sec2.1}]{Perini_2005,Perini_2005b,Perini_2012}. Also, images can function as a (description of a) scientific model, where the degree of similarity between model and the object it describes is a measure for its scientific \textquotedblleft success\textquotedblright\ \citep[][Section \ref{sec2.2}]{Downes_2009,Downes_2011,Downes_2012}.  Alternatively, Letitia Meynell argues that images can be effective tools to aid the visual imagination or image-ination and can provide insight when the topic of discussion is \textquotedblleft complex, surprising or unfamiliar\textquotedblright\ \citep[][Section \ref{sec2.3}]{Meynell_2017,Meynell_2018,Meynell_2020}. Meynell's account provides a plausible explanation for the function of the Stellar Graveyard: it may not be designed optimally for scientific analysis (e.g., to support an evidential claim or a useful model), but it is in fact a useful aide for the image-ination, helping readers visualize the observed black holes (Section \ref{sec2.4}).
    \item [\ref{sec3}.] Astrophysicists are tasked with explaining the cosmos, and in their method of \textquotedblleft saving the phenomena\textquotedblright\ the image-ination can be a useful tool to find explanations for observations (Section \ref{sec3.1}). Moreover, the image-ination can also be a tool to consider the qualitatively characteristic consequences of a certain theoretical scenario or alternative hypothetical scenarios, and engaging with the spatial structure of a system and the causal relationships within it can be seen as furthering scientific understanding \citep[][Section \ref{sec3.2}]{DeRegt_2017,Meynell_2020}. There is a variety of astrophysical imagery that appears to exhibit this interplay between image, imagination, and understanding (Section \ref{sec3.3}).
\end{itemize}
These findings, I argue, indeed show why the Stellar Graveyard is such an attractive image, despite its sub-optimal displaying of the binary black hole masses. That is, because the image shows the black holes as individual objects, it is easier for the reader to form mental images of these individual binary black hole orbiting each other and eventually merging. 
\begin{figure}
    \centering 
    \includegraphics[width=\linewidth]{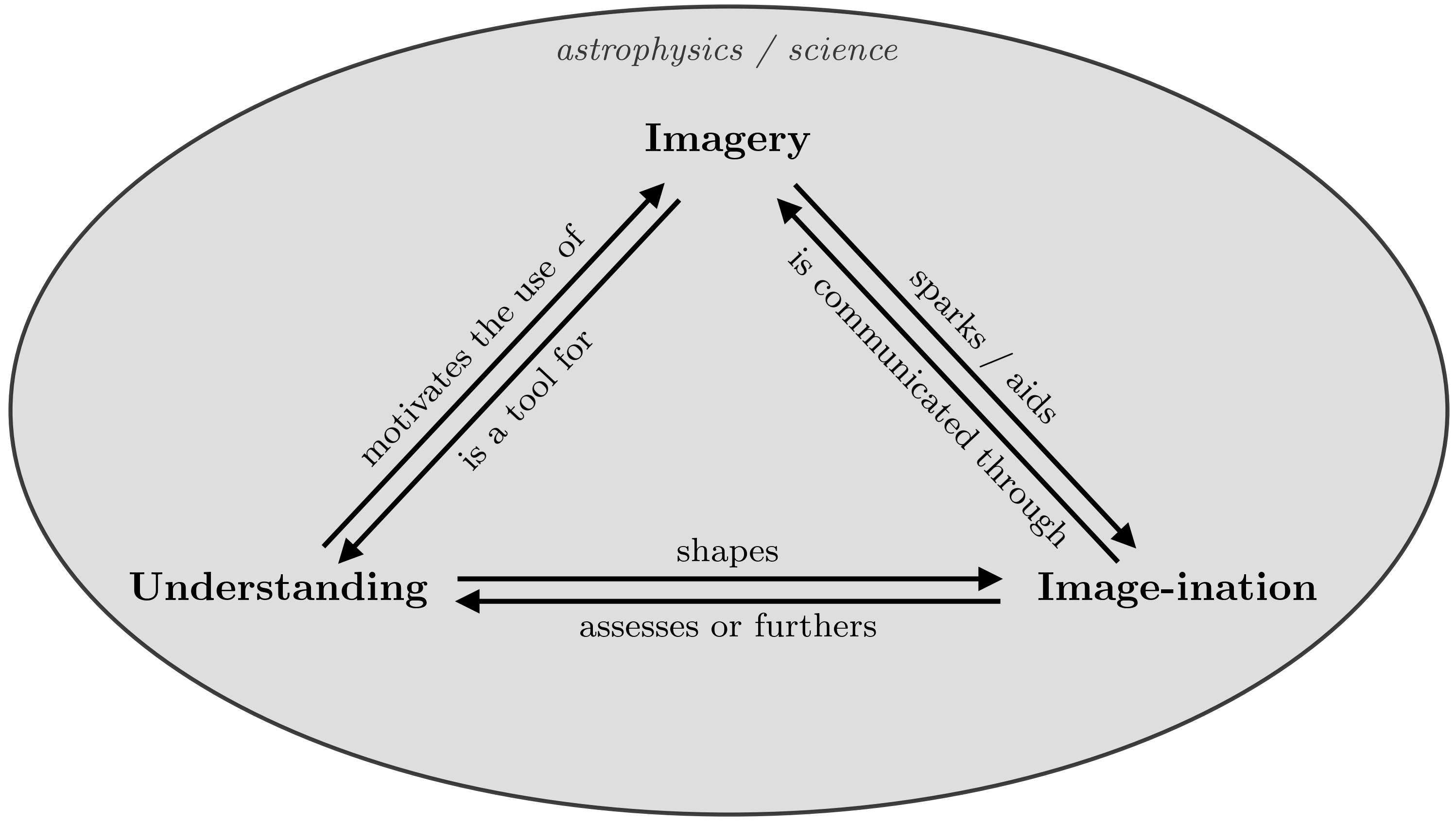}
    \caption{Diagram showing the relationship between the central concepts of this article: imagery, image-ination, and understanding, within the context of astrophysics / science.}
    \label{fig8}
\end{figure}
\par I have discussed the interaction between imagery, imagination, and understanding within the context of science (in particular, astrophysics). The interplay between these concepts is shown in Figure \ref{fig8}, and contains the following relations, based on the conclusions summarized above:
\begin{itemize}[label=\textbf{--}]
    \item Imagery can be used to \textquotedblleft spark,\textquotedblright\ or \textquotedblleft aid,\textquotedblright\ the image-ination. Mental images that are important to an author (e.g. in relation to understanding), can then be communicated to other scientists through actual imagery.
    \item Image-ination (i.e., the forming of mental images) can be used to assess the intelligibility of a theory, or even to further one's understanding of it (as it can, e.g., help in going over hypothetical scenarios). Also, one's image-ination is shaped by one's understanding of the topic, since understanding can entail grasping the relevant spatial structures and causal relationships.
    \item Understanding, as an important aim of science, motivates the use of the kind of images that reveal, for instance, spatial structures and causal relationships. After all, imagery can be used as a tool for scientific understanding, for example through visualizing qualitatively characteristic consequences.
\end{itemize}
In this article, I have focused on how images can interact with the image-ination and scientific understanding, particularly in the field of astrophysics. However, the main conclusions concerning the relationship between these three concepts are prevalent in, but certainly not limited to astrophysics. That is, in other fields of science images can, of course, be used in a similar way, and the scientific image-ination and understanding can work in similar ways across scientific disciplines. Because of this, I suspect that my findings regarding the interaction between imagery, image-ination, and understanding can be generalized and interpreted as a description of science in general. This is, of course, not addressed in this article but could be subject of future research.
\par As a final remark I turn to \citet{Downes_2012}, who briefly mentions that most scientific images are important for the argument presented in articles (in order to motivate his philosophical investigation of scientific imagery). In particular, he states:
\begin{quote}
    \textquotedblleft If you look through a random selection of current editions of science journals on the web, you will find thousands of pictures used in many and varied ways. Some could be dismissed as decorative or unnecessary but most, I argue, are indispensable to the relevant scientific work.\textquotedblright\ \citep[][p.\ 116]{Downes_2012}.
\end{quote}
Although I agree that most scientific images are indispensable for the scientific argument being made in the article, I don't think that philosophers should \textquotedblleft dismiss\textquotedblright\ images that are deemed \textquotedblleft decorative or unnecessary,\textquotedblright\ given the amount of care that goes into preparing scientific articles and the strict evaluations by editors and referees. After all, the astrophysical imagery discussed in the present article shows that Meynell's work provides a useful description of imagery that may fall into this category pf \textquotedblleft decorative or unnecessary\textquotedblright\ but still fulfills a genuine function in the communication between scientists, and this kind of imagery is therefore a valuable tool in describing scientific practice. That is, these astrophysical images should not be dismissed since they can provide valuable insight into the interplay between imagery, imagination, and understanding.

\newpage
\phantomsection
\addcontentsline{toc}{section}{References}
\spacing{1}


\setstretch{1.0}
\renewcommand{\bibsection}{\section*{\Large{References}}}
\bibliographystyle{chicago}
\bibliography{references_arXiv}

\end{document}